\documentclass[aps,pre,showpacs,groupedaddress]{revtex4}

\usepackage{graphicx}
\usepackage{fancyhdr}
\begin{document}


\title{Ordering dynamics of snow under isothermal conditions}


\author{H.~L\"{o}we}
\email[]{loewe@slf.ch}
\author{J.~K.~Spiegel}
\author{M.~Schneebeli}
\affiliation{WSL, Institute for Snow and Avalanche Research SLF, Davos, Switzerland}


\date{\today}

\begin{abstract}
  We have investigated the morphological evolution of laboratory new
  snow under isothermal conditions at different temperatures
  $T=-3,-9,-19\,^{\circ}C$ by means of X-ray tomography. The
  collective dynamics of the bicontinuous ice-vapor system is
  monitored by the evolution of the two-point (density) correlation
  function $C(\mathbf{r},t)$ and a particular thickness distribution
  which is similar to a pore size distribution. We observe the absence
  of dynamic scaling and reveal fundamentally different classes of
  length scales: The first class comprises the mean ice thickness and
  the (inverse) specific surface area (measured per ice volume) which
  increase monotonically and follow a power law. The dynamic exponent
  is in accordance with coarsening of a locally conserved order
  parameter. A second class of length scales is derived from the
  slopes of $C(\mathbf{r},t)$ at the origin in different coordinate
  directions. All these scales show a slower growth with anomalous
  power law scaling. The existence of two different power laws is
  quantitatively consistent with coarsening of fractal aggregates and
  reveal the persistence of power law correlations in the initial
  condition where the snow contains dendritic structures. At low
  temperatures these structures persist even over an entire year. A
  third class of length scales can be defined by the first zero
  crossings of $C(\mathbf{r},t)$. The zero crossings display a
  non-monotonic evolution with a strong anisotropy between the
  direction of gravity and horizontal directions. We attribute this
  behavior to larger scale structural relaxations of the ice network
  which apparently leave the small scale interfacial relaxations
  unaffected. However, vice versa it remains the question how
  structural mobility is induced by interfacial coarsening.
\end{abstract}


\maketitle

\section{Introduction}\label{sec:1}

Snow crystals exhibit a variety of morphological changes which are
driven by fundamentally different, thermodynamic conditions. During
growth the problem can be well described by an \emph{isolated} crystal
in a supersaturated environment which is a classical problem of single
crystal growth \cite{saito_1996}. If specifics of the ice crystal
lattice are taken into account, orientation dependent growth
velocities can be measured \cite{libbrecht_2003} and the snow
morphology diagram can be reasonably well explained in terms of
temperature, supersaturation and lattice properties, see
\cite{libbrecht_2005} for a review. If these non-equilibrium growth
forms are deposited (e.g. on the ground to form a seasonal snowpack)
the crystals no longer evolve in isolation. Structural correlations
may be built up already during atmospheric aggregation
\cite{westbrook_2004} or deposition on the ground \cite{loewe_2007}
where a tenuous ice network is formed. The subsequent evolution of the
deposit is then ultimately dominated by \emph{collective} behavior and
mutual interactions between the crystals. The simplest case of
thermodynamic conditions applied to a snow crystal deposit are
isothermal conditions at fixed temperature below zero. The subsequent
relaxation of the ice network to equilibrium is commonly referred to
as isothermal metamorphism of snow.

The occurrence of isothermal metamorphism in nature is limited to deep
polar snowpacks \cite{arnaud_2000}. From the collective dynamics point
of view the problem is however of general interest. Previous work
focused on possible transport mechanisms as the origin of interfacial
relaxation. It is widely believed that evaporation-condensation of
vapor is the dominant process of mass transport
\cite{domine_2003,legagneux_2004,legagneux_2005}. Experiments are then
interpreted in terms grain growth \cite{legagneux_2004}, sintering
theory \cite{kaempfer_2007} or a mean-field approach
\cite{legagneux_2005} which is similar to the LSW theory (Lifshitz and
Slyozov \cite{lifshitz_1961} and Wagner \cite{wagner_1961}) with
screening \cite{yao_1993}. For a review we refer to
\cite{blackford_2007}. The dynamical evolution of the ice morphology
is usually characterized in terms of a single characteristic length
scale $l$. Widely used is the inverse specific surface area per ice
volume
\cite{flin_2003,domine_2003,legagneux_2004,kaempfer_2007}. Sometimes
an ice thickness is used \cite{kaempfer_2007} which is derived from a
pore-size like thickness distribution. Another length scale observed
during isothermal metamorphism is the inverse mean curvature as
investigated in \cite{flin_2003,flin_2004}. However, there is no
agreement about the dynamic exponents $z$ which governs the growth law
$l\sim t^{1/z}$. Recent Monte Carlo simulations \cite{vetter_2009}
confirm power law behavior between $2$ and $4$ with a strong variation
with temperature. Some work suggest a logarithmic growth $l\sim
\ln(t)$ \cite{legagneux_2004} at intermediate times. Most of the work
investigates natural snow samples which resist to reveal a concise
picture of the dynamics.

To address the values of the exponents and a possible origin of the
scattering it is instructive to first resort to certain idealized
pictures with universal features as a reference. Given the
evaporation-condensation mechanism, snow as an ice-vapor mixture in a
non-equilibrium state will relax to equilibrium by capillarity driven
coarsening \cite{ratke_2002}. Neglecting gravity for a moment, it is
widely accepted \cite{bray_1994} that basically two different
universality classes are likely to occur in phase ordering systems:
For a locally conserved order parameter, all length scales $l$ in the
system will evolve with $z=3$ whereas the locally non-conserved case
leads to $z=2$. Note, that the specification of
evaporation-condensation as the dominant transport mechanism alone
does not ultimately determine the dynamic exponent: Depending on
whether the dynamics is limited by intermediate diffusion or the
interface reaction the order parameter is effectively locally or
globally conserved, respectively. Both cases are treated in LS
\cite{lifshitz_1961} whereas W only considers the globally conserved
case \cite{wagner_1961}. The picture is however neither limited to
isolated droplet-like structures nor to infinite dilution as in
LSW. Both universality classes are also found in simulations of the
late stage evolution of (symmetric) bicontinuous structures as
exemplified by the Cahn--Hilliard or Allen--Cahn equation
\cite{kwon_2007}. In both cases \emph{dynamic scaling} scaling holds
and the system is truly characterized by a single diverging length
scale. Within this idealized picture snow should end up with $z=2$ or
$z=3$. Additionally any of the commonly investigated quantities such
as specific surface area, mean or Gaussian curvature can be equally
likely used as the characteristic length scale to determine the
dynamic exponent $z$.

Indeed, real systems almost always display deviations from
ideality. This is a consequence of emerging, competing length
scales. The origin of additional scales can roughly be distinguished
into the following classes: i) transient behavior from initial
conditions which involve intrinsic non-short ranged correlations, ii)
the emergence of intermediate dynamic regimes which are governed by
different scaling laws. Both lead to a an apparent breakdown of
dynamic scaling under experimental conditions. For snow it is likely
that both effects will play a role. The most unambiguous measure to
detect the presence (or absence) of dynamic scaling are the rescaling
properties of length distribution functions. Amongst others the most
prominent candidate is the equal time, two-point (or pair) correlation
function $C(\mathbf{r},t)$ of the microscopic density
$\rho(\mathbf{x},t)$. Its importance stems from the fact that its
Fourier transform, the structure factor is directly accessible by
scattering experiments of inhomogeneous media \cite{debye_1949}
without explict knowledge of the microscopic density
$\rho(\mathbf{x},t)$. In addition two-point correlation functions are
a general starting point for systematic approaches to physical
properties of heterogeneous materials \cite{torquato_2002}.

The focus of the present paper is a multiple-scale characterization of
the isothermal evolution of snow from its collective dynamics point of
view, which is probed by fluctuations of the microscopic density field
$\rho(\mathbf{x},t)$. We use X-ray tomography to measure
$\rho(\mathbf{x},t)$ with high resolution throughout an entire year
for different temperatures and characterize the dynamics by means of
the density correlation function. Surprisingly, this has not been
investigated before. To better control the initial conditions we use
laboratory snow with crystals grown from vapor which are subsequently
deposited into sample holders by sieving. By comparing the evolution
of various length scales we are able to identify different mechanisms
which contribute to the breakdown of dynamic scaling. These mechanisms
are likely to be active simultaneously in natural snow samples.

The paper is organized as follows. The experimental setup is described
in Section \ref{sec:2}. In Section \ref{sec:3} we define the
correlation function and a thickness distribution from which various
length scales are derived. The measurement results for the observables
are presented in Section \ref{sec:4}. The discussion of all results is
done Section \ref{sec:5} and in the final section \ref{sec:6} we give
the conclusions and suggestions for future work.

\section{Experiments}\label{sec:2}

Our general experimental setup for the X-ray tomography basically
follows \cite{kaempfer_2007}, with two important differences for the
sample handling: i) the isothermal storage was improved to minimize
the effect of predominant temperature gradients, ii)
laboratory-generated new snow is used to guarantee similar initial
conditions for the snow samples at different temperatures. Details are
given below.

\subsection{Snow sample preparation}

\begin{figure}[t]
  \begin{center}
  \includegraphics[width=0.49\textwidth]{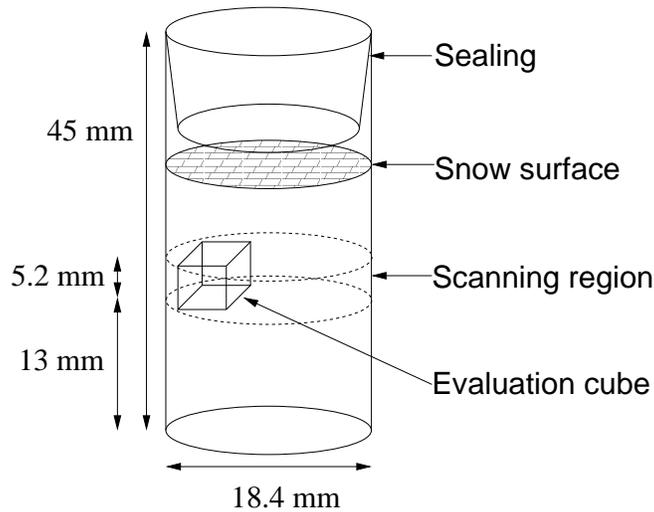}%
  \end{center}
  \caption{\label{fig:1} Schematic of the cylindrical tomography sample
    holder.}
\end{figure}

All snow samples were prepared from ``nature-identical'' laboratory
snow which was produced in a simple snowmaker as described below. The
design basically follows \cite{nakamura_1978}: A heated water
reservoir is kept at warm temperature of $T=30\,^{\circ}C$ in a cold
room environment at ambient room temperature of
$T=-25\,^{\circ}C$. The humid air above the water surface is
continuously advected (by a fan) into box at ambient temperature where
the vapor precipitates at thin nylon wires which serve as growth
nuclei. By varying air and water temperature the method is able to
qualitatively reproduce growth modes and crystal habits predicted by
the snowflake morphology diagram \cite{libbrecht_2005}. Under the
conditions described above, mainly dendritic structures are
generated. Snow crystals are periodically harvested from the wires by
automatic vibrations of the wires. The so obtained snow powder compact
had a density of $\rho\approx 172\,\mathrm{kg\,m^{-3}}$. With an ice
density of $\rho_{\mathrm{ice}}=917\,\mathrm{kg\,m}^{-3}$ this amounts
to initial volume fractions of $\phi_{\rm i}=0.188$. The snow was
stored for 24\,h at $T=-25\,^{\circ}C$ to allow for moderate, initial
sintering of the crystals. Using a sieve with mesh size of 2 mm, the
snow was sieved into cylindrical sample holders, cf.~Fig~\ref{fig:1},
until they were completely filled. The snow was then slightly
compressed and sealed with a polyethylene cap (PE-LD) to avoid
sublimation and obtain a closed environment with respect to mass
exchange.

\subsection{Sample storage}
To improve the isothermal conditions, the storage boxes used in
\cite{kaempfer_2007} were improved to reduce thermal gradients. The
storage consists of layers of highly insulating and highly conducting
materials. The highly conducting material with high heat capacity
(steel) prevents the buildup of a temperature inversion in the air,
and eliminates temperature fluctuations. From the inside out, the
packaging was as follows:
The sample cylinders were fixed in a Styrofoam mask inside a steel
cylinder with 1\,cm thick walls. The steel cylinders were capped from
both sides with 1\,cm thick steel lids. This steel cover was packed
into a small Styrofoam box with 5\,cm thick walls. The box was again
surrounded by 5\,mm thick steel plates and then enclosed in a vacuum
box.  The temperature was recorded inside each steel cylinder by two
temperature sensors (iButtons DS1922L). These sensors have an accuracy
of $\pm 0.5\,^{\circ}C$, (from $-10\,^{\circ}C$, to $+65\,^{\circ}C$,)
and a resolution of $0.0625\,^{\circ}C$. They were calibrated before
the measurements.  Recorded temperatures were almost constant with
mean values and standard deviations of $T=-3.7\pm 0.7\,^{\circ}C$,
(sample 1), $T=-9.5\pm 0.1\,^{\circ}C$, (sample 2)and $T=-18.6\pm
0.8\,^{\circ}C$, (sample 3). This corresponds to a homologous
temperature of 0.99, 0.97 and 0.93.

\subsection{Tomography}
At the beginning of the experiment each sample is measured roughly in
3 week intervals. Since the evolution slows down during the experiment
\cite{kaempfer_2007} we decreased the measuring frequency in the late
stage of the experiment. For measurements the samples are transported
from the storage room to the CT in a small Styrofoam box. We use a
commercial $\mu$-CT80 micro-computer-tomograph (CT) from Scanco.  The
scanning in the CT was done with a nominal resolution (pixel size) of
10\,$\mu$m and a modulation transfer function at 10\% contrast level
of 12.4\,$\mu$m. The scanning region was fixed in space at 13\,mm
height from the bottom of the sample holder, cf.~Fig.\ref{fig:1}. The
height of the scanning region is 5.2\,mm (520 voxels). The CT was kept
close to the nominal storage temperatures -3, -9 and
$-19\,^{\circ}C$. After the measurement, the sample was placed in the
transport box and returned to the storage box.

From the attenuation map of the scanned region a cube of $520^3$
voxels = $(5.2\,\mathrm{mm})^3$ was extracted, cf.~Fig.~\ref{fig:1},
which was large enough to be a representative volume for the
considered properties \cite{kaempfer_2005}. The gray scale images were
filtered using a Gaussian filter with kernel-size of $5^3$ and
standard deviation of 1.2 voxels to improve the signal to noise
ratio. A binary image was then obtained by segmentation of the
filtered data. A single segmentation threshold was used for all
images. The threshold was determined such that the density of all
samples is matched on average.

\section{Observables}\label{sec:3}

\subsection{Two point correlation function}

Appropriate measures to monitor the evolution of the microstructure
can be defined from the microscopic density. For continuous mass
distributions the microscopic density is defined in terms of the phase
indicator function $\phi_{\mathrm{i}}(\mathbf{x})$ of the ice phase
which is defined by $\phi_{\mathrm{i}}(\mathbf{x})=1$, if $\mathbf{x}$
lies in the ice phase and zero otherwise. The mass density is then
related by the intrinsic density of ice
$\rho_{\mathrm{ice}}=917\,\mathrm{kg\,m}^{-3}$ via
\begin{equation}
  \label{eq:defrho}
  \rho_{\mathrm{i}}(\mathbf{x},t):=\rho_{\mathrm{ice}}\,\phi_{\mathrm{i}}(\mathbf{x},t)\,.
\end{equation}
The simplest first order statistical quantity of a two phase random
medium is its volume fraction
\begin{equation}
  \label{eq:defphi}
  \phi_{\mathrm{i}}(t):=\overline{\phi_{\mathrm{i}}(\mathbf{x},t)}\,.
\end{equation}
The overbar denotes ensemble averaging. Practically we replace
ensemble averages by volume averages and thus implicitly assume a
statistically homogeneous system.

The simplest, higher order statistical characterization of a random,
two-phase medium is given by the equal-time, two-point correlation
function
\begin{equation}
  \label{eq:defC}
  C(\mathbf{r},t):=\overline{
    \big[ \phi_{\mathrm{i}}(\mathbf{x}+\mathbf{r},t)-\phi_{\mathrm{i}}(t) \big]
    \big[ \phi_{\mathrm{i}}(\mathbf{x},t)-\phi_{\mathrm{i}}(t) \big]}\,.
\end{equation}
The two-point function $C(\mathbf{r},t)$ is the continuous counterpart
of the pair correlation function of discrete particle assemblies.
Note that $C(\mathbf{r},t)$ does not only depend on the magnitude of
$\mathbf{r}$, thus we explicitly account for anisotropic
behavior. Below we will restrict ourselves on the behavior along
different coordinate directions and therefore define
$C_{\alpha}(r,t):=C(r\,\mathbf{e}_{\alpha},t)$ for $\alpha=x,y,z$.

From the two-point correlation function various length scales can be
defined. For later purposes we follow \cite{lipshtat_2002} and define
a length scale from the slope of $C(\mathbf{r},t)/C(0,t)$ at the
origin, viz.
\begin{equation}
  \label{eq:defl}
  l_{\alpha}(t):=- \left(\frac{d}{dr} C_{\alpha}(r,t)/C_{\alpha}(0,t)\Big|_{r=0}\right)^{-1}\,.
\end{equation}
We note that for isotropic media in spatial dimension $d=3$ the
correlation lengths are related to the specific surface area $s$ per
\emph{unit volume} via $\frac{d}{dr} C_{\alpha}(r,t)\Big|_{r=0}=s/4$
for $\alpha=x,y,z$, cf. \cite{torquato_2002}. Henceforth the
$l_{\alpha}$ are referred to as interfacial (correlation) lengths.

In bicontinuous materials, such as microemulsions, the covariance is
often approximated by oscillatory two-scale forms,
e.g. $C(r)/C(0)=\exp(-r/l)\,\cos(\pi r/(2l_0))$ which involves,
besides the interfacial correlation length $l$ another length scale
$l_0$. The latter determines the first zero crossing of $C(r)$ which
is attained at $r=l_0$ and can be given the meaning of a typical
domain size. In view of this two-scale approximation we define the
first zero crossing
\begin{equation}
  \label{eq:defl_0}
  l_{0,\alpha}(t):=\min\{r|C_{\alpha}(r,t)=0\}
\end{equation}
as another characteristic length scale, which is referred to the
structural (correlation) length henceforth.

Finally, we define the ratio of ice volume and surface area as an
additional length scale since it is widely used in snow science,
\begin{equation}
  \label{eq:defl_ssa}
  l_{\mathrm{ssa}}(t):=\phi_{i}(t)/s(t)\,.
\end{equation}

\subsection{Thickness distribution}
Another commonly used distribution of length scales for a porous
material has been defined in \cite{hildebrand_1997} and is given by
the distribution of local thicknesses. Its definition deviates from
the common pore-size distribution \cite{torquato_2002} and can be
written as a conditional probability
\begin{equation}
  \label{eq:defp_th}
  p_{\mathrm{th}}(r,t):=
  \overline{ \delta\big( r-r_{\mathrm{max}}(\mathbf{x}) \big)
    \,\phi_{\mathrm{i}}(\mathbf{x},t)}
  /\phi_{\mathrm{i}}(t)\,,
\end{equation}
where $r_{\mathrm{max}}(\mathbf{x})$ is the radius of the largest
sphere which i) contains $\mathbf{x}$ (not necessarily as its center)
and ii) is itself contained completely in the ice phase. An important
implication of this definition of thickness is revealed by the
following example. Consider a collection of non-overlapping spheres of
radius $r_0$. This indeed implies a delta-like thickness distribution
$p_{\mathrm{th}}(r)=\delta(r-r_0)$. If the spheres are reshaped (under
volume conservation) to a long cylinder of radius $r_0$ which is
terminated by hemi-spherical caps of the same radius at both ends the
thickness distribution remains invariant.

As a characteristic length
scale of the thickness distribution we use the mean which is given by
\begin{equation}
  \label{eq:defl_th}
  l_{\mathrm{th}}(t):=\int_0^{\infty} dr\;r\,p_{\mathrm{th}}(r,t)\,.
\end{equation}

\section{Results}\label{sec:4}

\subsection{Overview}

\begin{figure}[t]
  \includegraphics[width=0.49\textwidth]{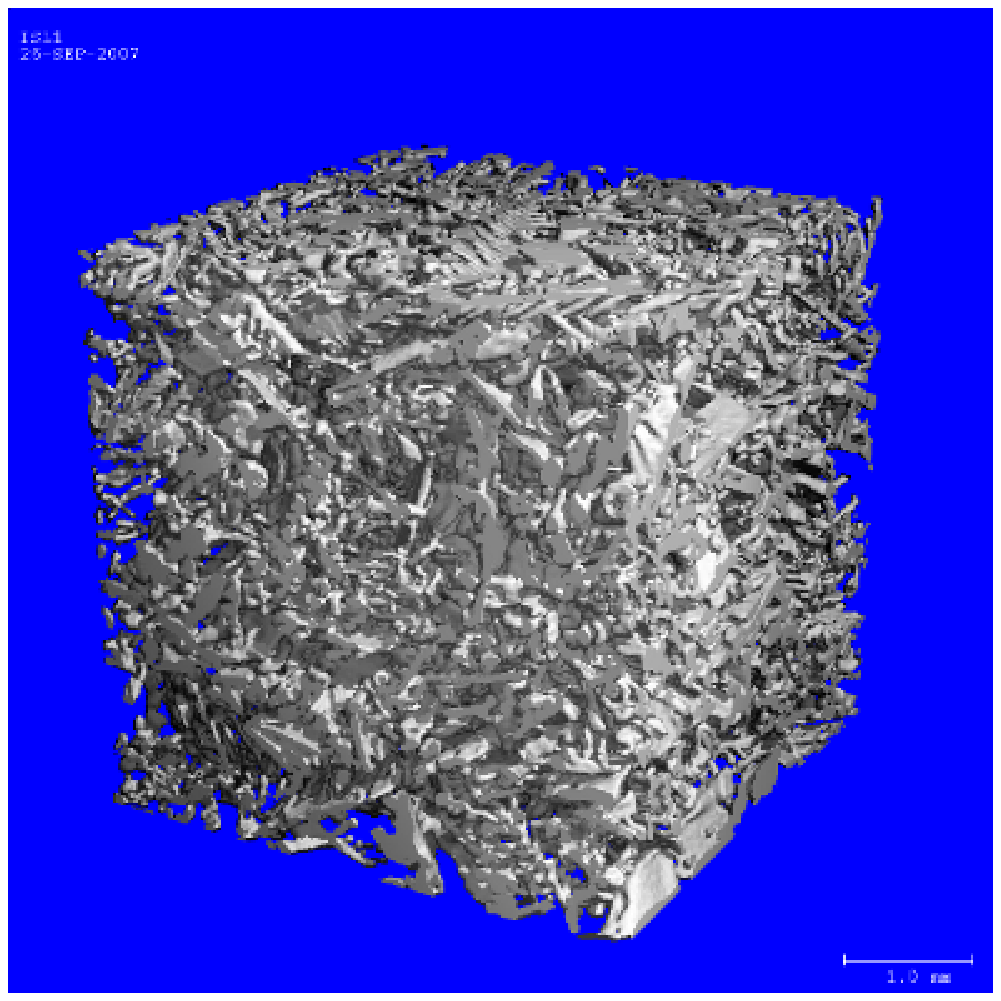}%
  \hfill
  \includegraphics[width=0.49\textwidth]{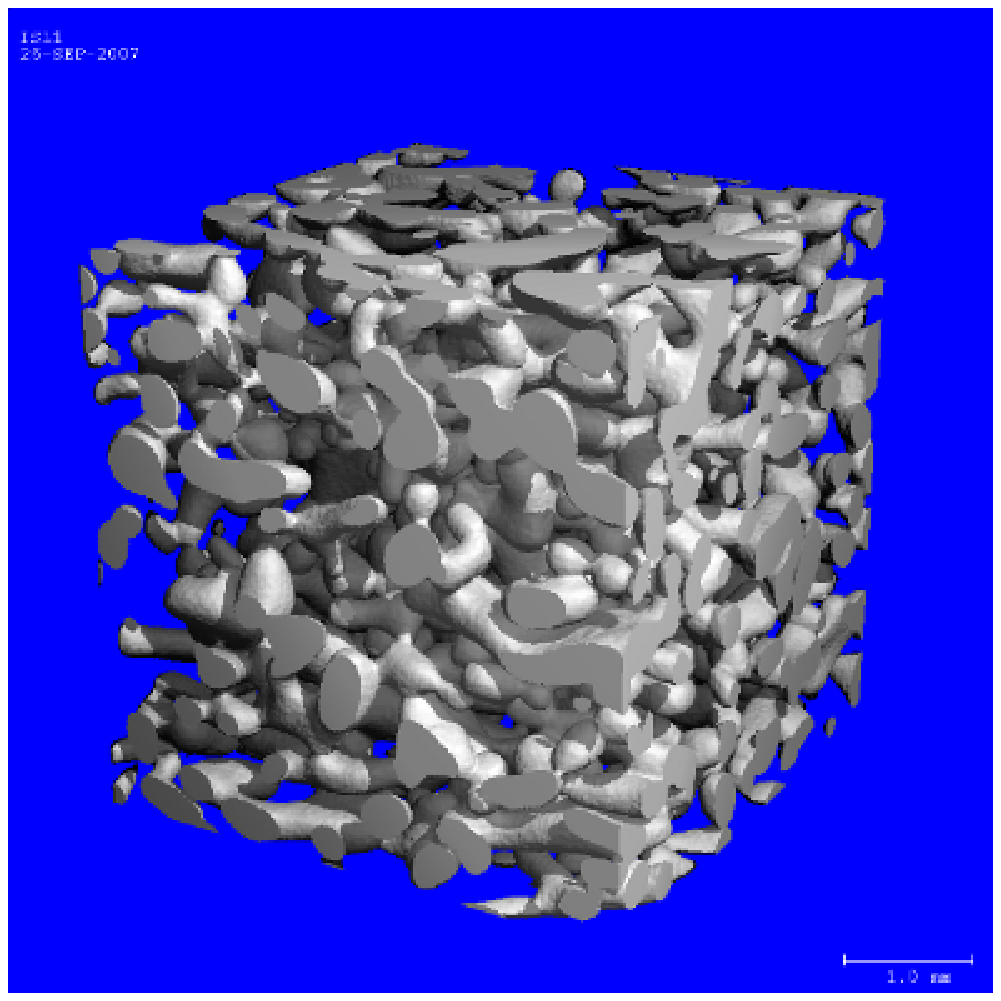}%
  \caption{\label{fig:3} (Colors online) Evolution of a snow cube at
    $T=-3\,^{\circ}C$, in the beginning (left), and in the end after
    almost one year (right). The length of the cube is 5.2 mm.}
\end{figure}
The structural analysis of the snow samples is evaluated within the
cubic subsets of the entire sample of size $(5.2\,\mathrm{mm})^3$,
cf.~Fig.~\ref{fig:1}.  With a voxel size of 10\,$\mu$m this leads to
lattice sizes of $520^3$ . A visual impression of the three
dimensional evolution of the ice-matrix is shown in
Fig.~\ref{fig:3}. The samples evolve from highly branched, fresh snow
in the initial conditions to a rounded bicontinuous structure. Warmer
samples evolve faster than colder ones.

\subsection{Density}
\begin{figure}
  \includegraphics[width=0.49\textwidth]{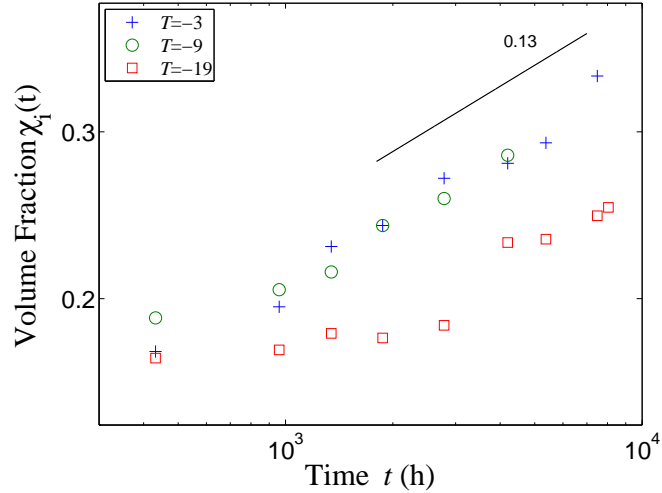}%
  \caption{\label{fig:4} (Colors online) Temporal evolution of the
    volume fraction $\phi_{\mathrm{i}}(t)$ at all temperatures.}
\end{figure}

The evolution of the volume fraction given in Eq.~(\ref{eq:defphi}) of
the samples is shown in Fig.~\ref{fig:4}. Note that the scale is
double logarithmic and a guide to the eye is given as a reference for
a possible power-law evolution at the late stage. However, the
evolution clearly displays deviations from a straight line caused by
modulations.

\subsection{Two point correlation function}\label{sec:Results}

\subsubsection{Scaling}
\begin{figure}
  \includegraphics[width=0.49\textwidth]{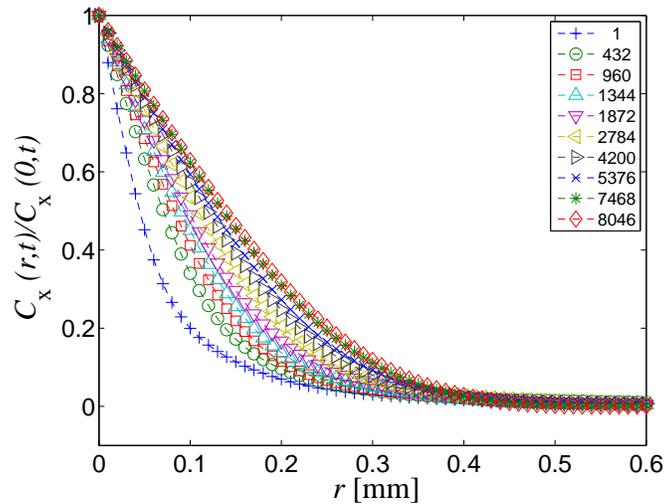}%
  \caption{\label{fig:5} (Colors online) Evolution of $C_x(r,t)$ at
    $T=-3\,^{\circ}C$ for different times (in h). On a linear scale
    the evolution at all temperatures appears to be similar.}
\end{figure}

In the following we follow \cite{lipshtat_2002} and normalize the
correlation function defined in Eq.~(\ref{eq:defC}) by their values at
the origin $C_{\alpha}(r,t)/C_{\alpha}(0,t)$. An overview of the
evolution of the two-point correlation function $C_x(r,t)$ at
$T=-3\,^{\circ}C$ is given in Fig.~\ref{fig:5}. All samples clearly
display a cusp singularity at the origin revealing a non-fractal
ice-vapor interface on scales $10\mu$m throughout the entire
experiment. The smooth appearance of the interface on these scales has
also been confirmed by a comparison between X-ray tomography and gas
absorption (BET) derived surface area \cite{kerbrat_2008}.

Next we consider the, possibly anisotropic, dynamic scaling and
rescale $C_{\alpha}(r,t)$ by the correlation lengths $l_{\alpha}(t)$
which are obtained by fitting the cusp at the origin against
$C_{\alpha}(r,t)/C_{\alpha}(0,t)=1-r/l_{\alpha}(t)$. These slopes
increase in the course of time revealing a coarsening of the
structure. From the perspective of dynamic scaling the results at
$T=-9\,^{\circ}C$ cannot be distinguished from $T=-3\,^{\circ}C$ and
thus we concentrate on the comparison between $T=-3\,^{\circ}C$ and
$T=-19\,^{\circ}C$. In addition, at both temperatures the $x$ and $y$
directions can be regarded as equivalent and thus we concentrate on
the comparison between $x$ and $z$ direction. We explicitly note that
$z$ is the direction of gravity.
\begin{figure}
  \includegraphics[width=0.49\textwidth]{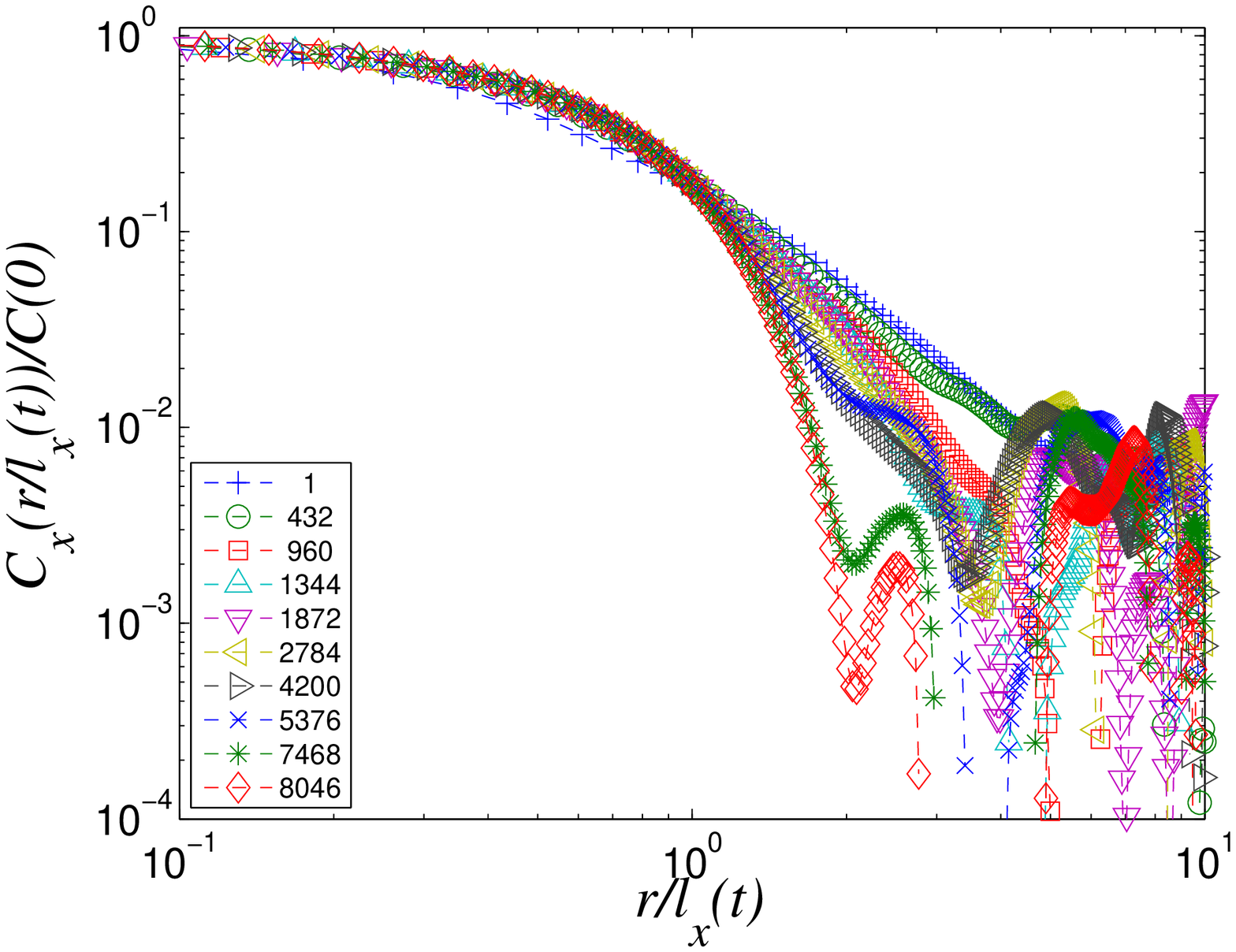}%
  \hfill
  \includegraphics[width=0.49\textwidth]{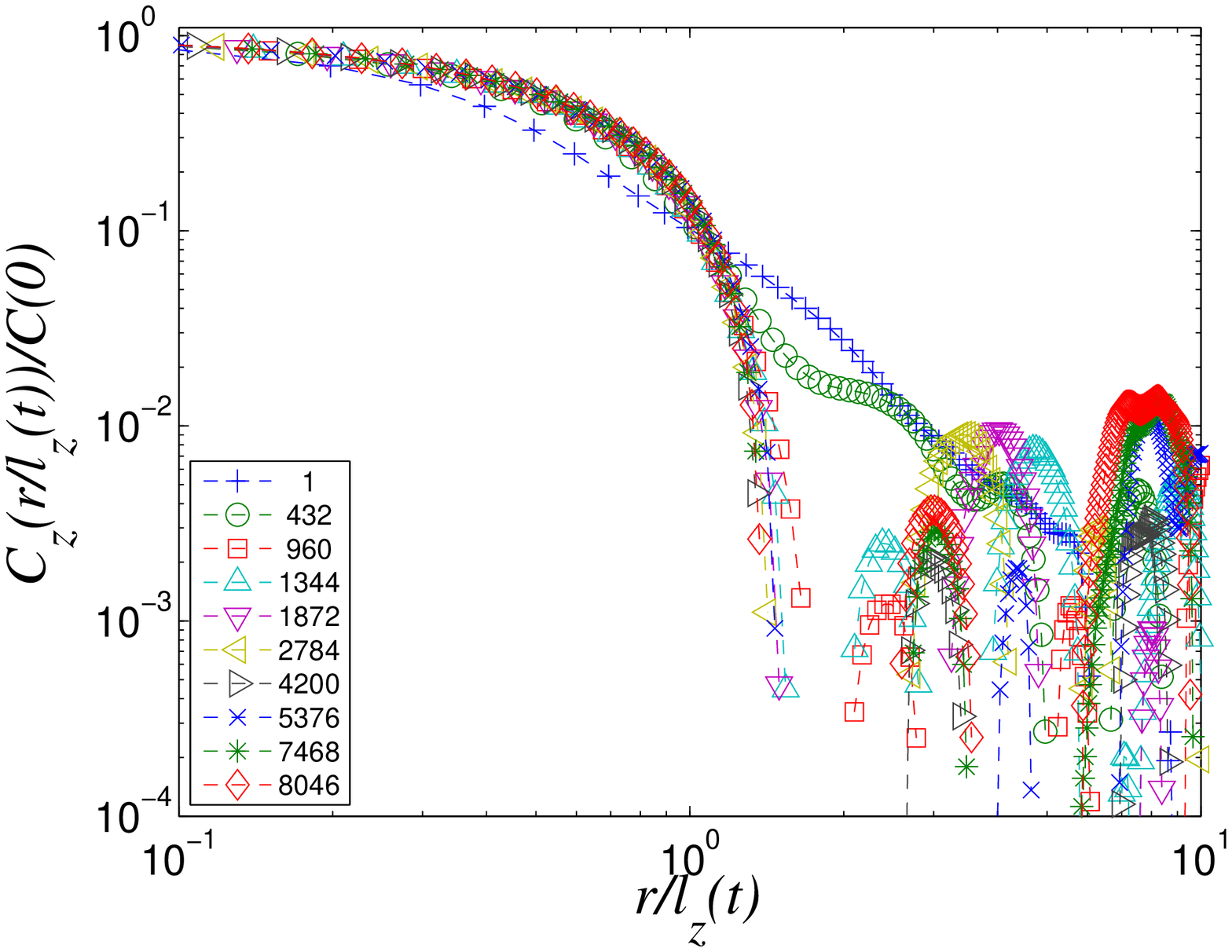}%
  \caption{\label{fig:6} (Colors online) Rescaled two-point
    correlation functions for $T=-3\,^{\circ}C$. Left: horizontal
    direction $C_{\mathrm{x}}$ Right: vertical direction
    $C_{\mathrm{z}}$ . The behavior at $T=-9\,^{\circ}C$ is similar.}
\end{figure}

\begin{figure}
  \begin{center}
  \includegraphics[width=0.49\textwidth]{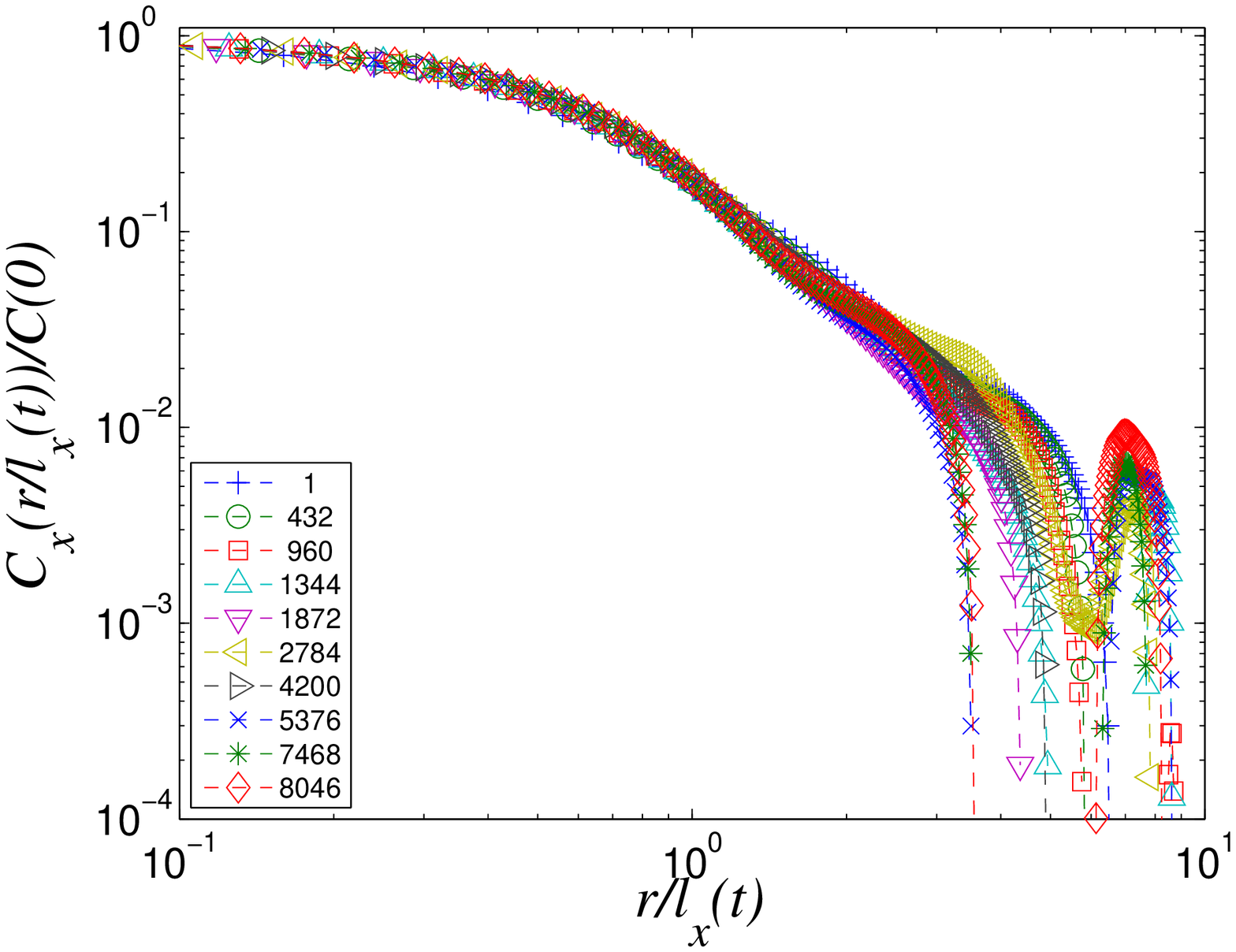}%
  \hfill
  \includegraphics[width=0.49\textwidth]{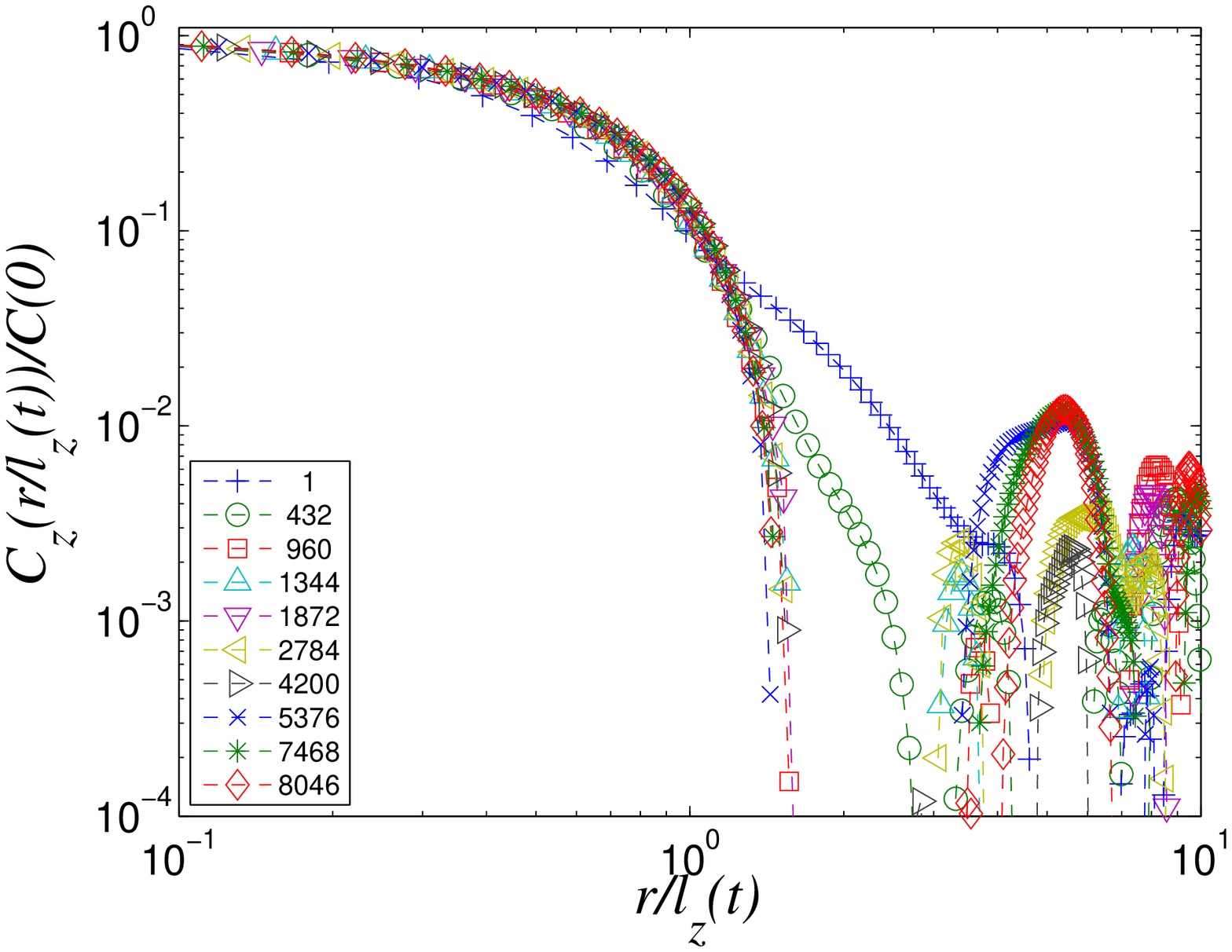}%
  \end{center}

  \caption{\label{fig:7} (Colors online) Rescaled two-point
    correlation functions for $T=-19\,^{\circ}C$. Left: horizontal
    direction $C_{\mathrm{x}}$ Right: vertical direction
    $C_{\mathrm{z}}$ .}
\end{figure}
The results for the rescaled correlation functions at high temperature
$T=-3\,^{\circ}C$ are given in Fig.~\ref{fig:6}, and for the low
temperature $T=-19\,^{\circ}C$ in Fig.~\ref{fig:7}. All plots are
shown on a double logarithmic scale since the quality of the data
collapse on larger scales remains usually unrevealed on a linear
scale. For both temperatures a reasonable data collapse in $z$
direction is already established after a short time and extends beyond
the first zero crossing which is above $r/l_{\alpha}(t)\approx 1$. In
contrast in $x$ direction the scaling is poor and extends hardly
beyond $r/l_{\alpha}(t)\approx 1$. Only at the very late stage after
almost one year the quality of the scaling is comparable to that in
$z$ direction. In any case, the rescaling behavior of $C$ is poor,
indicating the absence of dynamic scaling and the existence of
different relevant length scales.

\subsubsection{Initial condition and final state}

\begin{figure}[t]
  \begin{center}
  \includegraphics[width=0.49\textwidth]{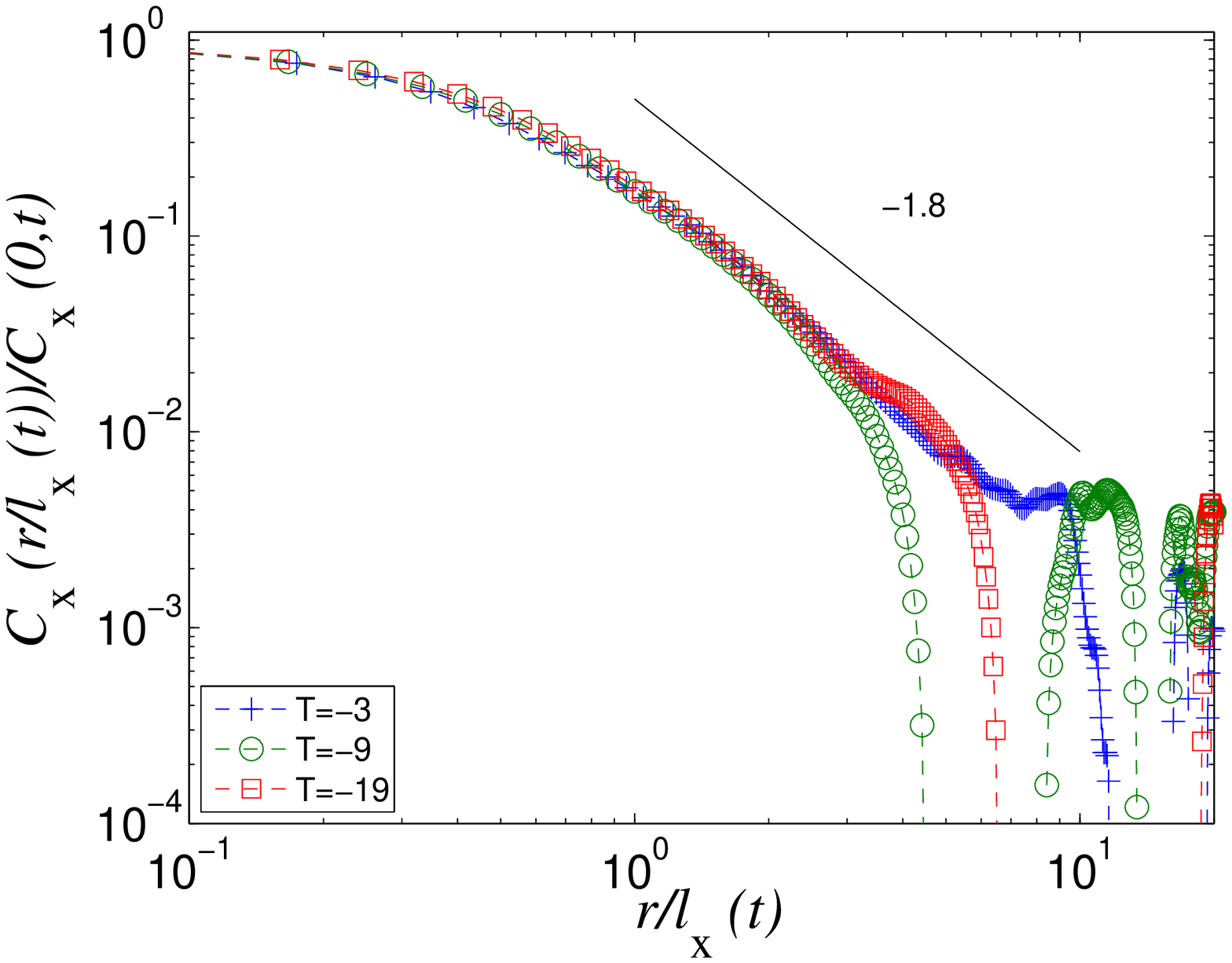}%
  \hfill
  \includegraphics[width=0.49\textwidth]{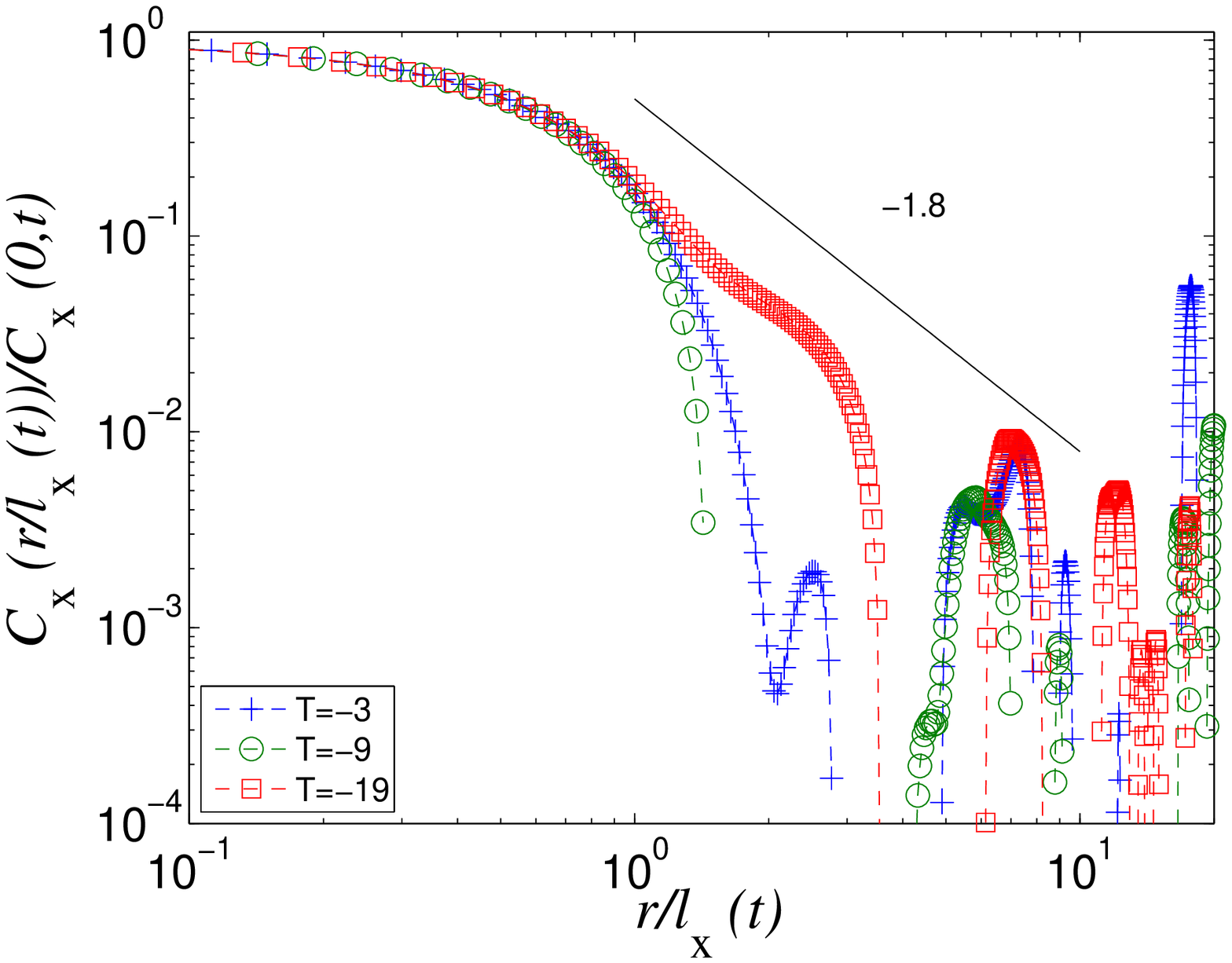}%
  \end{center}
  \caption{\label{fig:8} (Colors online) Rescaled correlation
    functions of the initial (left) and the final state (right) for
    all temperatures.}
\end{figure}
To shed additional light on the initial condition and the final states
obtained after the entire year of coarsening the correlation functions
for both states are compared for all temperatures in
Fig.~\ref{fig:8}. Initially, correlations clearly exist beyond
$r/l_{\alpha}(t)\approx 1$ in all three coordinate directions. The
decay can approximately described by a power-law with an exponent
$C(r/l_{\alpha}(t))\sim (r/l_{\alpha}(t))^{-\beta}$ with
$\beta=1.8$. This behavior is not very striking and extends only over
a range of one magnitude. However, it clearly indicates the presence
of correlations on scales large compared to $l_{\alpha}(t)$. The
origin of this order is revealed by a closer, visual inspection of the
sample,
\begin{figure}[t]
  \begin{center}
  \includegraphics[width=0.45\textwidth]{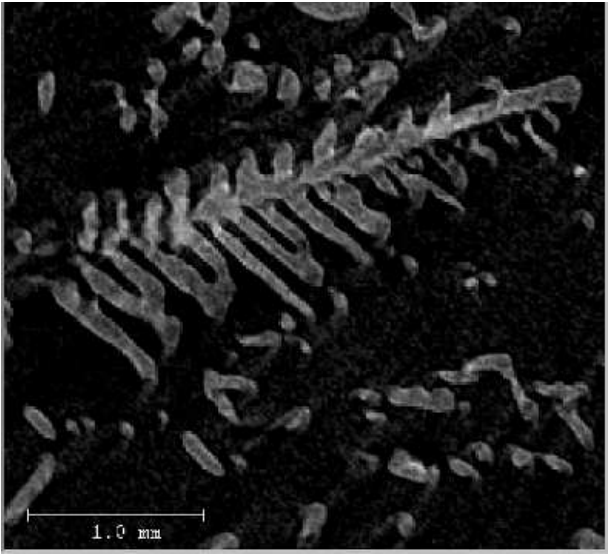}%
  \hfill
  \includegraphics[width=0.45\textwidth]{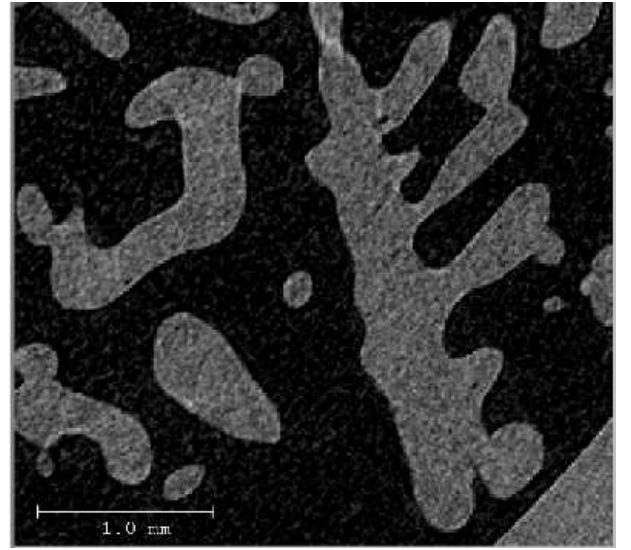}%
  \end{center}
  \caption{\label{fig:9} Dendritic structures in snow at
    $T=-19\,^{\circ}C$ in the initial condition (left) and after one
    year (right) found in the $x-y$ plane. Note that the structures
    are not identical.}
\end{figure}
which reveals the presence of dendritic structures,
cf.~Fig.~\ref{fig:9}. For the lowest temperature the structures
persist throughout the entire year. Again by (subjective) visual
inspection we confirm that after one year at $T=-19\,^{\circ}C$
dendrites can only be found in the $x-y$ plane, cf.~Fig.~\ref{fig:9}.
whereas we could not find dendritic structures from views in the $x-z$
or $y-z$ plane. This is in agreement with the behavior found for the
correlation function in Fig.~\ref{fig:7}: in $x$ direction (also in
$y$ direction, not shown) at $T=-19\,^{\circ}C$ the initial power-law
is still visible in form of a knee-like feature prior to the first zero
crossing of the correlation function. In contrast in $z$ direction any
signature of the initial power-law correlations immediately
disappears.

Before investigating the time evolution of the correlation lengths in
detail we shall consider the distribution of ice thicknesses.

\subsection{Ice thickness distribution}

\subsubsection{Scaling}
The ice-thickness distribution (\ref{eq:defp_th}) is computed from the
vendor software of the CT device by a distance transform of the ice
phase \cite{hildebrand_1997}. The evolution of the rescaled (by the
mean
\begin{figure}[t]
  \begin{center}
  \includegraphics[width=0.49\textwidth]{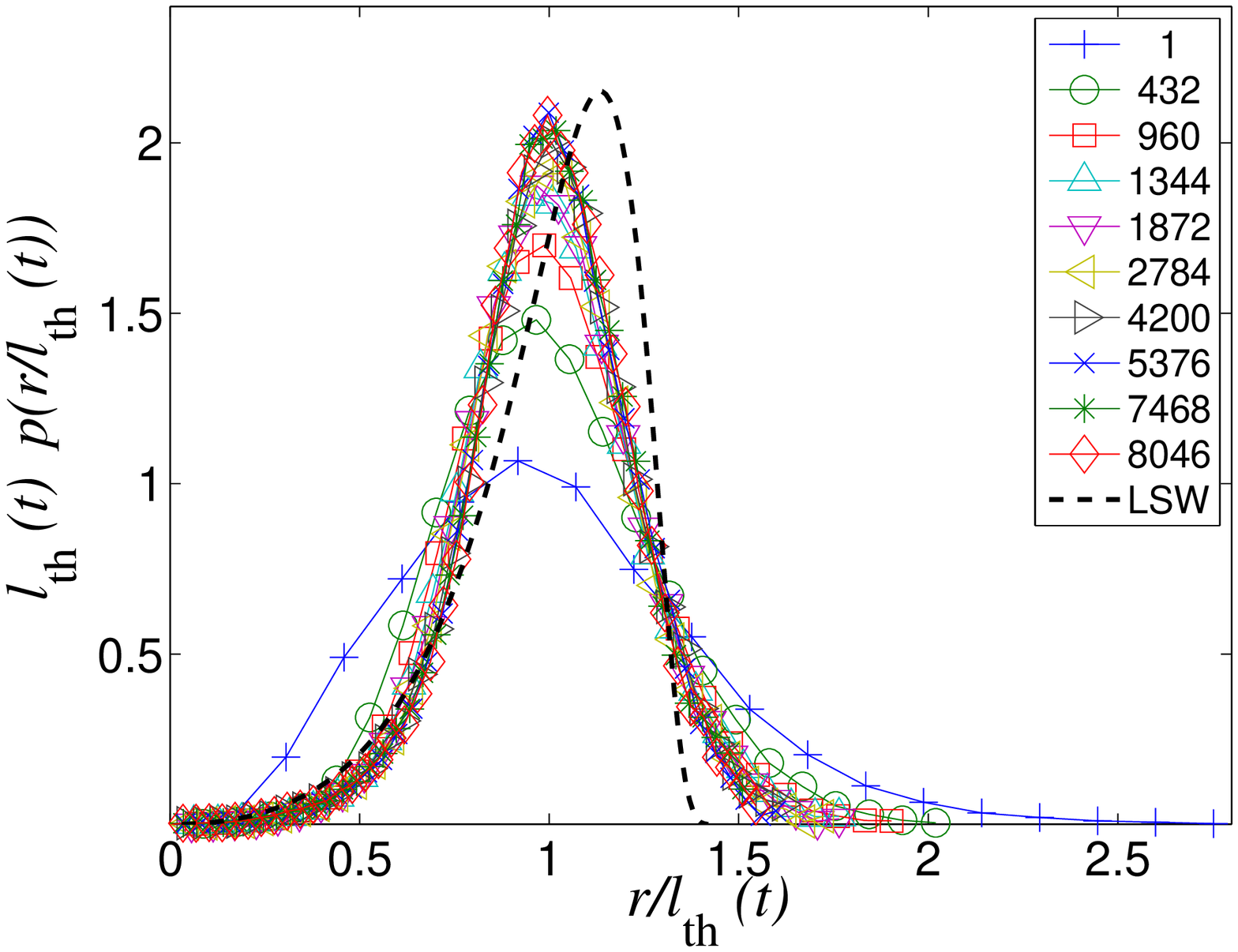}%
  \hfill
  \includegraphics[width=0.49\textwidth]{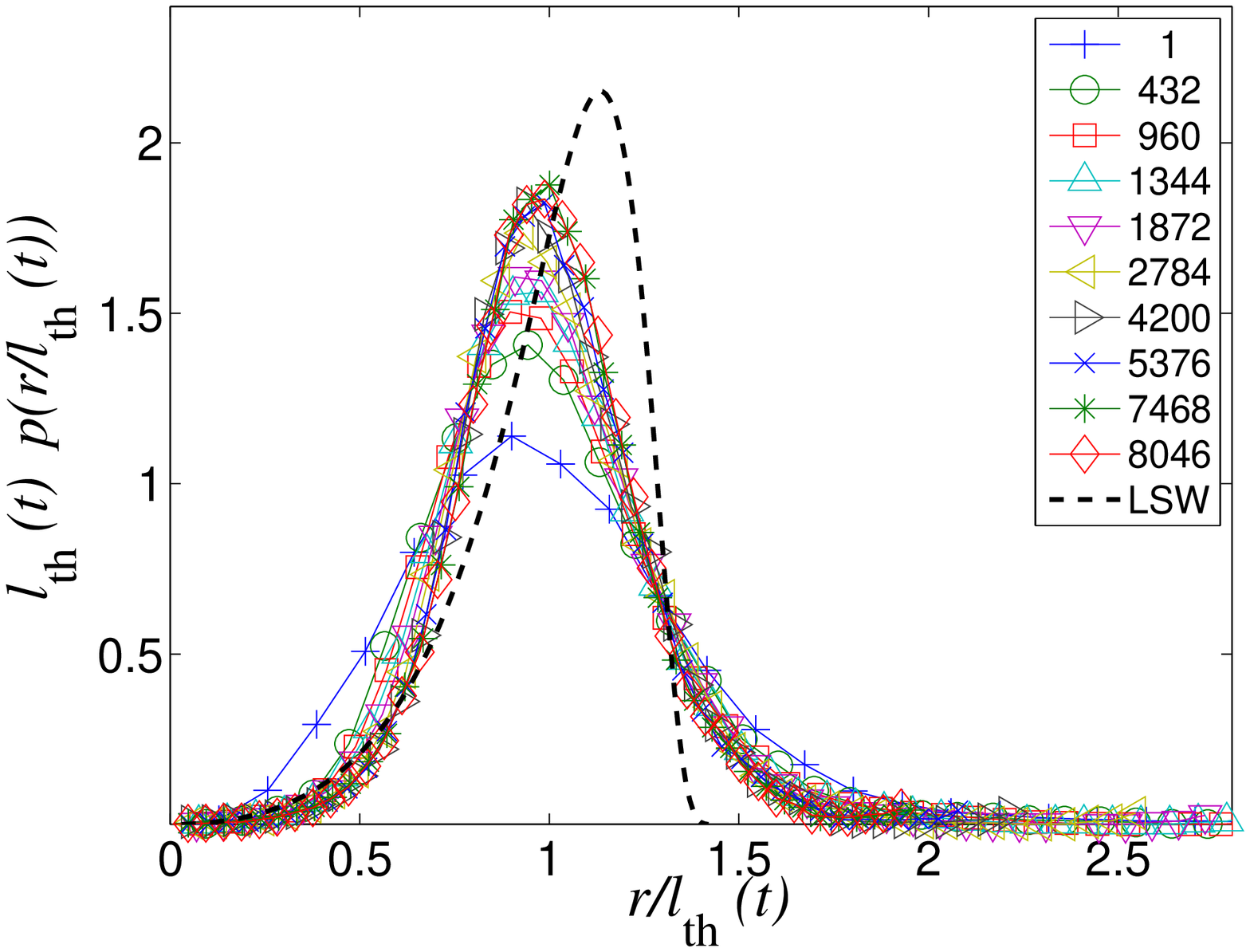}%
  \end{center}
  \caption{\label{fig:10} (Colors online) Rescaled ice thickness
    distribution functions for $T=-3\,^{\circ}C$ (left), and
    $T=-19\,^{\circ}C$ (right).}
\end{figure}
$l_\mathrm{th}$) thickness distributions is given in
Fig.~\ref{fig:10}.  Since the thickness is a measure for an
approximation of the structure by spheres, we also compare the
thickness distribution to the classical radius distribution from
Lifshitz and Slyozov \cite{lifshitz_1961} which is additionally
plotted in Fig.~\ref{fig:10}. Again we concentrate on the comparison
between $T=-3,-19\,^{\circ}C$.

\subsubsection{Initial and final state}

The ice-thickness distribution of the initial and final states are
shown in Fig.~\ref{fig:7} on a double logarithmic scale.
\begin{figure}[t]
  \begin{center}
  \includegraphics[width=0.49\textwidth]{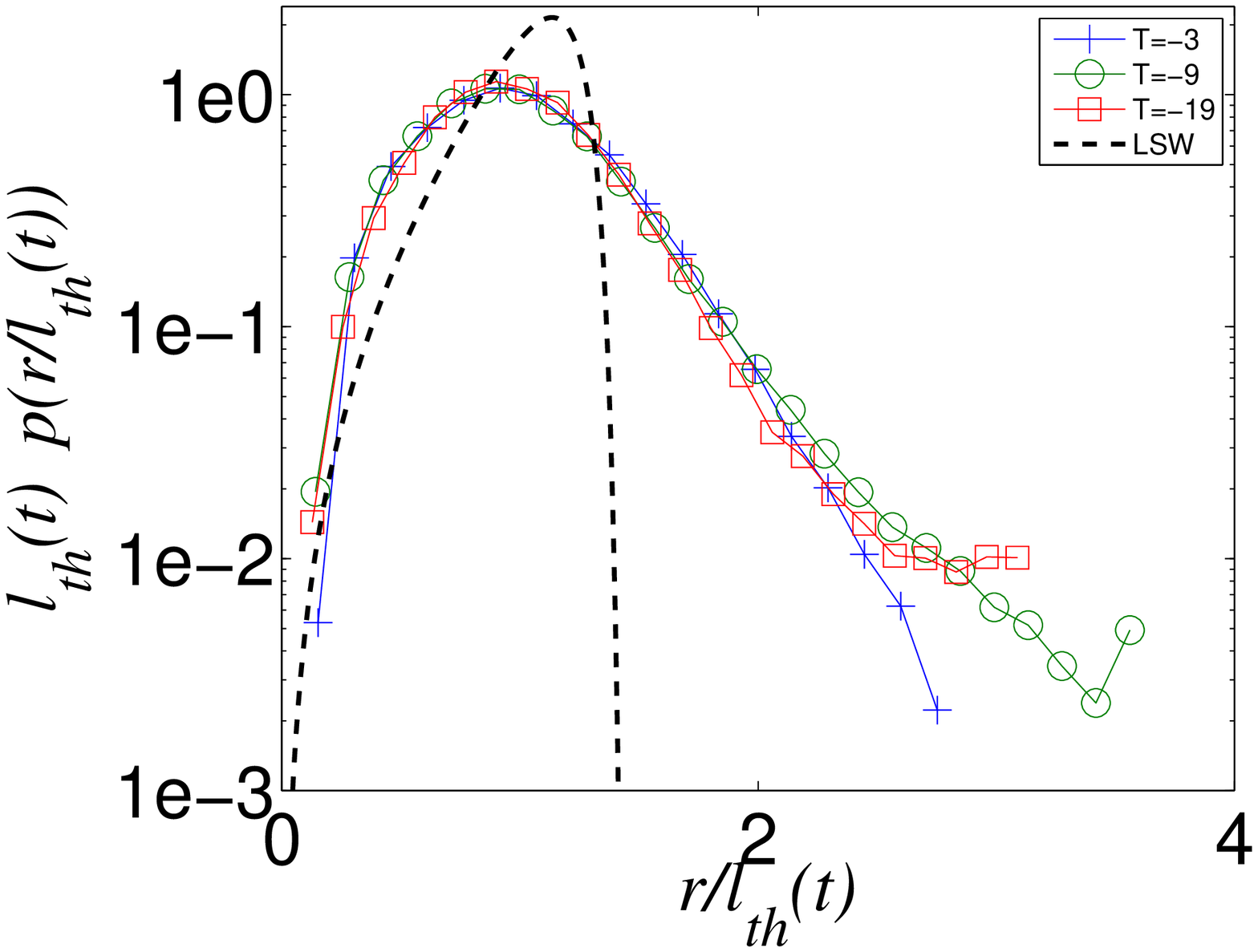}%
  \hfill
  \includegraphics[width=0.49\textwidth]{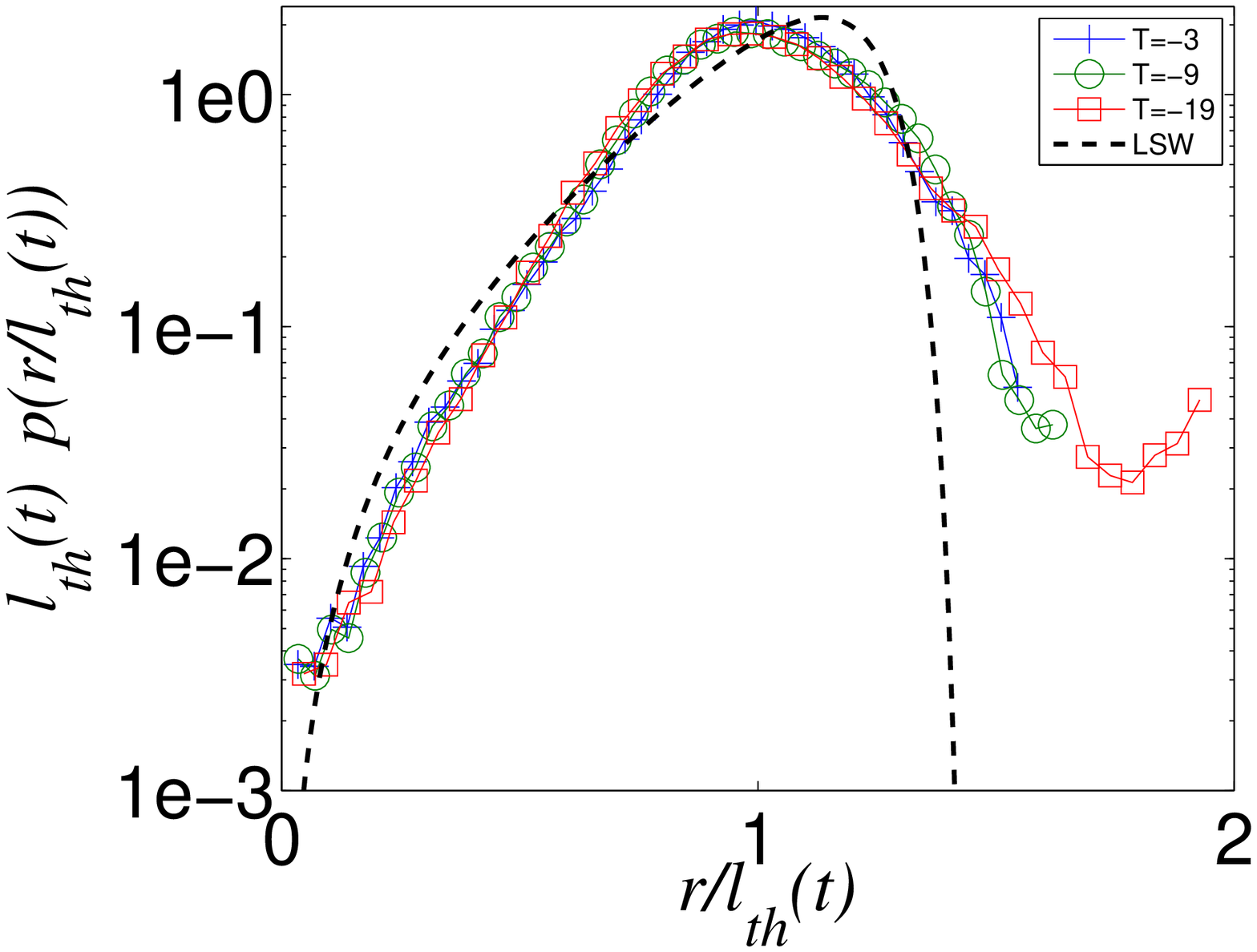}%
  \end{center}

  \caption{\label{fig:11} (Colors online) Rescaled ice thickness
    distribution of the initial (left) and the final state(right).}
\end{figure}
It is difficult to reveal the existing correlations beyond
$r/l_{\alpha}(t)\approx 1$ from the dendrites in the thickness
distribution. This is due to its incapability of detecting the largest
axis of anisotropic filaments. Therefore the interpretation of the
tails of the distribution is somewhat subtle. Since
$p_{\mathrm{th}}(r,t)$ is an isotropic measure, the emerging
anisotropy detected by the correlation function in different
coordinate directions also remains unrevealed.

\subsection{Length scales}\label{sec:lengthscales}

%
Finally we compare the time evolution of all length scales defined in
Sec.~\ref{sec:3} which are derived from $C(r,t)$ and $p_{\mathrm
  th}(r,t)$.

First we compare the interfacial correlation lengths $l_{\alpha}$, the
mean thickness $l_{\mathrm{th}}$ and the (inverse) specific surface
area per \emph{ice} volume $l_{\mathrm{ssa}}$. For a better comparison
we normalize all lengths by their initial values since absolute values
of growth rates are not of interest in the present study.
\begin{figure}
  \begin{center}
  \includegraphics[width=0.49\textwidth]{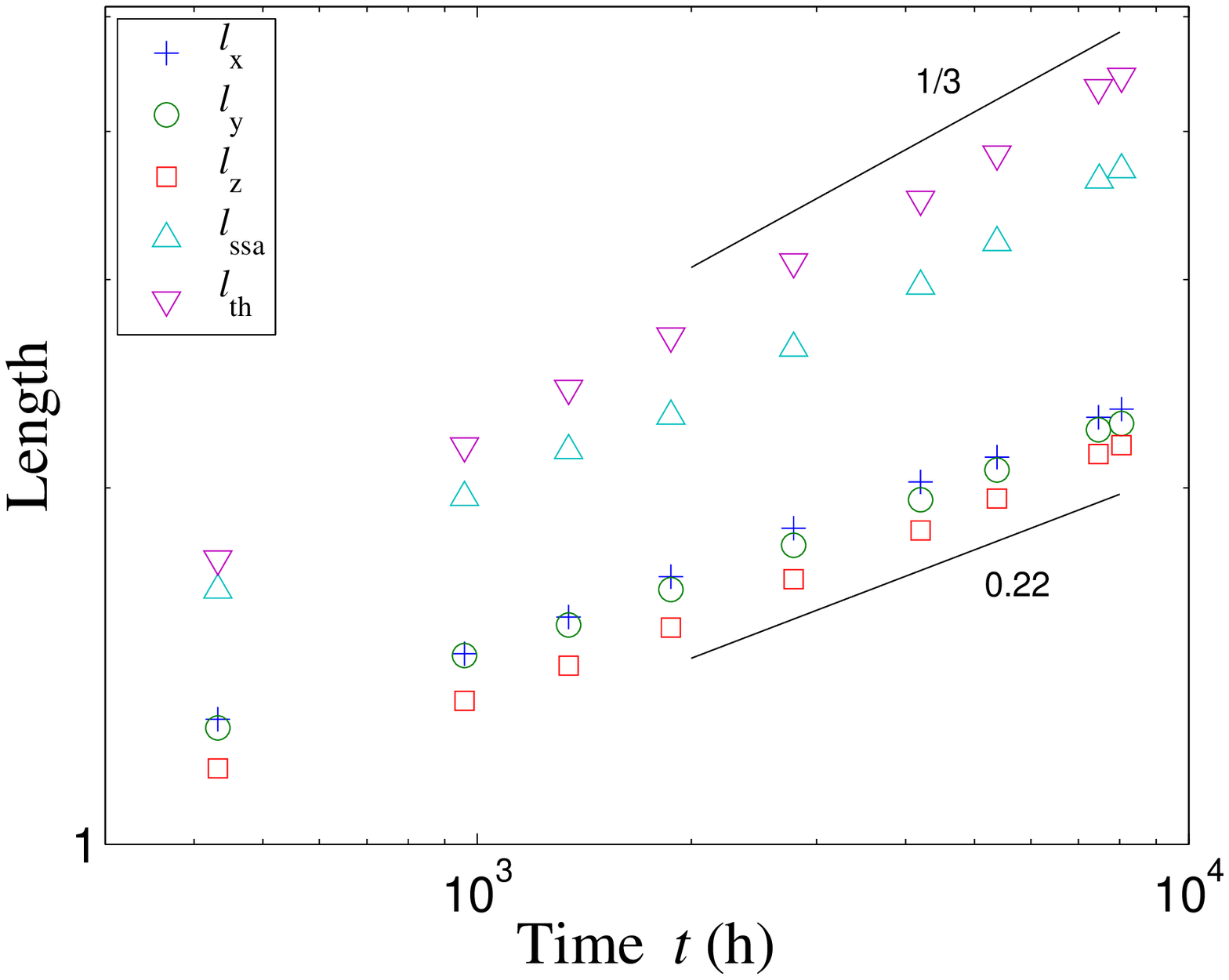}%
  \hfill
  \includegraphics[width=0.49\textwidth]{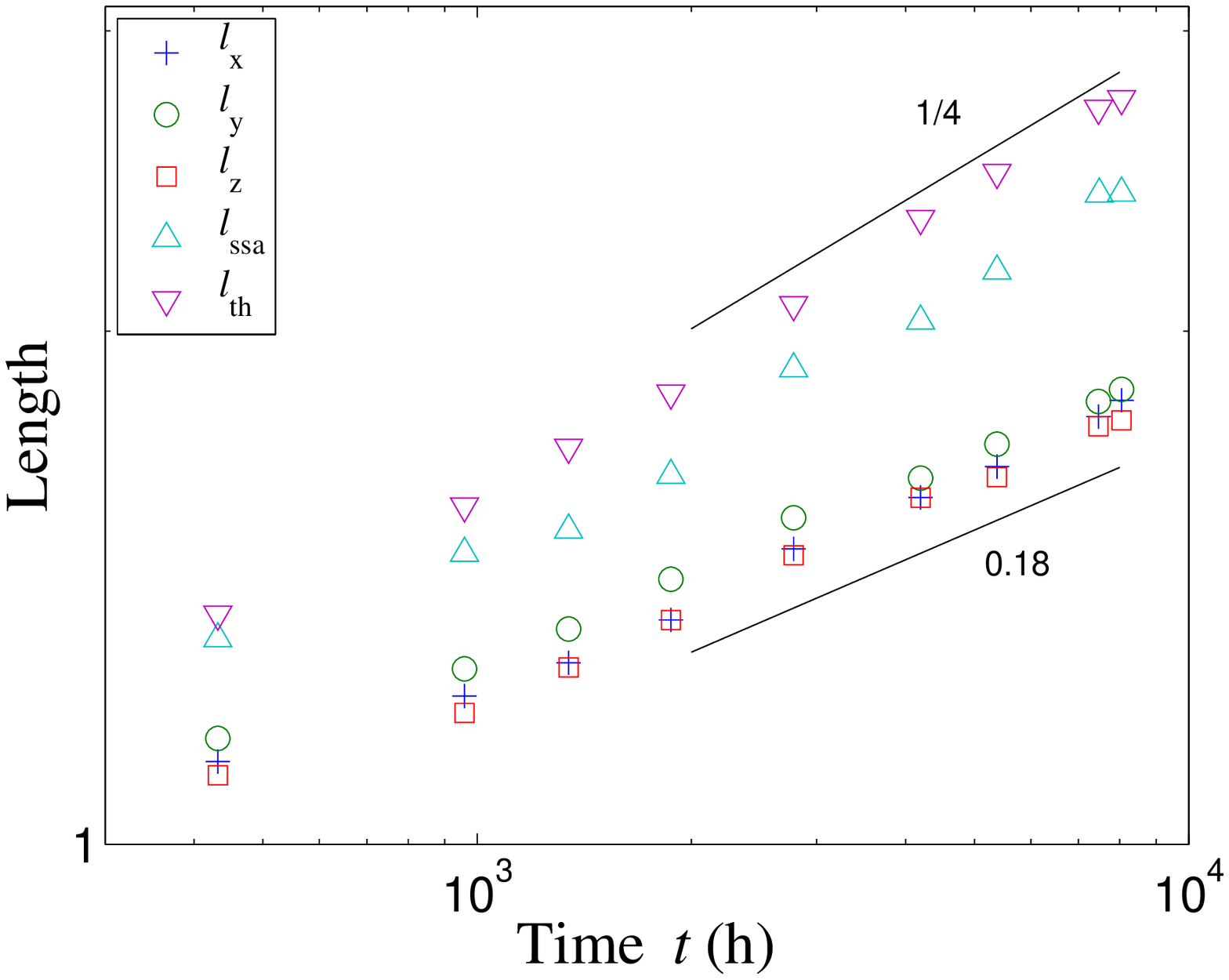}%
  \end{center}

  \caption{\label{fig:12} (Colors online) Time evolution of the
    interfacial correlation lengths $l_{\mathrm{th}}$ , the ice
    thickness $l_{\mathrm{ssa}}$ , and the (inverse) specific surface
    area $l_{\mathrm{ssa}}$ at $T=-3\,^{\circ}C$ (left) and
    $T=-19\,^{\circ}C$ (right).}
\end{figure}
The various length scales are shown in Fig.~\ref{fig:12}. All of them
show reasonable power law behavior with two different exponents which
are given as straight black lines as a guide to the eye.

\begin{figure}
  \begin{center}
  \includegraphics[width=0.49\textwidth]{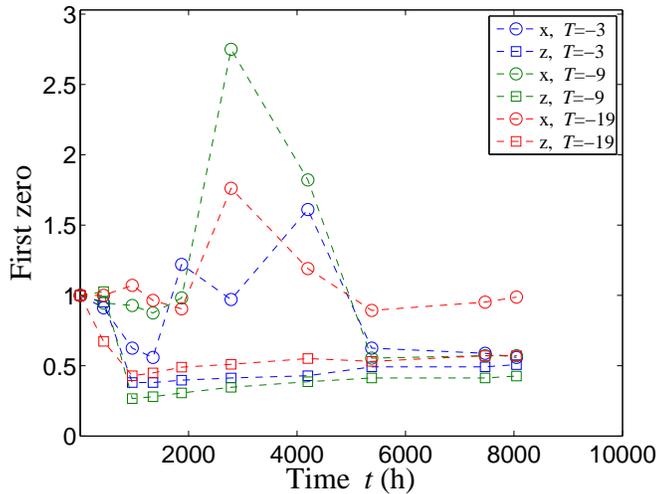}%
  \end{center}
  \caption{\label{fig:13} (Colors online) Time evolution of the
    structural correlation lengths $l_{0,\alpha}$ for $\alpha=x$
    (circles) and $\alpha=z$ (squares) at all temperatures.}
\end{figure}

Finally, we evaluate the first zeros $l_{0,\alpha}$ of the correlation
function in $x$ and $z$ direction in Fig.~\ref{fig:13}, again
normalized by their initial values.
Interestingly, a completely different, non-monotonic behavior is
observed: The first zero of the $C_{z}(r,t)/C_{z}(0,t)$ always
decreases in the beginning reaches a minimum after two time steps and
increases subsequently. The time at minimum coincides with the
beginning of the onset of data collapse of the $z-$correlation
function $C_{z}(r,t)/C_{z}(0,t)$ observed in
Fig.~\ref{fig:6},\ref{fig:7}. In contrast, the first zeros of
$C_{x}(r,t)/C_{x}(0,t)$ always display transient behavior in the
beginning which is difficult to characterize. After attaining a
maximum at intermediate times all scales seem to follow the slow
growth of the $z$ scales at the very late stage where all zeros grow
in unison.

\section{Discussion}\label{sec:5}

Our starting point of the discussion is the usual assumption of
dynamic scaling during isotropic coarsening \cite{bray_1994} which
manifests itself in a scaling form
\begin{equation}
  C_{\alpha}(r,t) = f(r/l(t)) 
\end{equation}
for the correlation function at long times independent of the
direction $\alpha$. Our measured correlation functions clearly reveals
the breakdown of dynamic scaling during isothermal metamorphism of
snow. This indicates a coupling between several length scales which
has to be elucidated.

Our first result is the existence of different classes of length
scales which can all be described by a power law, however with
different exponents, cf.~Fig.~\ref{fig:12}. This is precisely the
behavior reported in \cite{lipshtat_2002} for the coarsening of
fractal clusters as exemplified by the late stage dynamics of the
Cahn--Hilliard equation. The key ingredient is the presence of power
law correlations in the initial conditions which gives rise to an
additional relevant length scale during coarsening. In our case the
initial conditions contain dendritic structures which survive the
mechanical fragmentation process (sieving) during sample
preparation. These structures are revealed by signatures of a power
law tail in the the correlation function well above the interfacial
correlation length which is set by $r/l_{\alpha}(t)\approx 1$,
cf.~Fig.\ref{fig:6}. The decay can roughly be described by $C(r,0)\sim
r^{-1.8}$, cf.~Fig.~\ref{fig:8}. These correlations survive in form of
the knee-like feature in the correlation function
(cf.~Fig.~\ref{fig:6},\ref{fig:7}) over a large extent of time, as
predicted in \cite{lipshtat_2002}. Visually the small scale dynamic in
Fig.~\ref{fig:6} is pinned at the initially immobile large-scale
features (arms of the dendrites) which slows down the growth of small
scales. Only in $z$ direction the knee-like feature disappears almost
immediately after two time steps and the correlation function exhibits
reasonable data collapse up to its second zero crossing.

At high temperatures ($T=-3\,^{\circ}C$) the observed exponent values
in Fig.~\ref{fig:12} are close to $1/z=0.33$, $1/z=0.22$. The smaller
value for $l_{\alpha}(t)$ is in agreement with that obtained in
\cite{lipshtat_2002}. We note, that our definition of $l_{\alpha}(t)$
coincides with that given in their work, though our definition of the
correlation function differs. The authors in \cite{lipshtat_2002}
consider the two-point correlation function of a single, fractal
cluster with respect to the origin which decays to zero for large
distances. However, we are dealing with a collection of those clusters
and a system which is homogeneous on large scales. Accordingly the
two-point correlation function exhibits a non-zero value at large
arguments, and therefore we have defined $C(r)$ to be the
\emph{covariance} of the random microstructure which tends to zero for
large arguments. The larger of our power-law exponents is close to the
LS value $z=3$. We follow \cite{lipshtat_2002} and attribute this
behavior to the signatures of locally conserved order parameter
dynamics which is apparently detectable only in some length scales.

At low temperatures, cf.~Fig.~\ref{fig:12}, the scenario is
qualitatively identical to the high temperature case, only the values
of the exponents have changed indicating a different underlying
dynamics. Here, $l_{th}$ and $l_{ssa}$ follow an evolution which is
close to $z=1/4$ whereas the exponent for the interfacial correlation
lengths $l_{\alpha}$ are again well below this value. A dynamic
exponent $z=1/4$ is characteristic for surface diffusion
\cite{balluffi_2005}. This is consistent with the observations in
\cite{libbrecht_2003} where surface diffusion on ice is effectively
damped above $T=-15\,^{\circ}C$. On general grounds one may expect
however, that surface diffusion is only an intermediate dynamical
regime since the $l\sim t^{1/3}$ scaling of evaporation-condensation
will eventually always dominate the $l\sim t^{1/4}$ of surface
diffusion. Thus it is likely that even at low temperatures a terminal
LS value of $z=3$ is attained at even later times. These time scales
are however difficult to achieve experimentally. In addition, it is
reasonable to expect that the crossover time scale where initial
conditions of the surface correlation length have died out and the
intermediate surface diffusion or the terminal
evaporation-condensation dynamics is attained will itself increase
with decreasing temperature. Thus, effective (time dependent) dynamic
exponents $z(t):=d\,\ln(l)/d\,\ln(t)$ are measured,
cf. \cite{huse_1986} and an apparently continuous increase of the
measured exponents with decreasing temperature is likely to occur. In
analogy, to the high temperature case we attribute the exponent $z$
which is obtained from $l_{th}$ and $l_{ssa}$ to the true dynamic
behavior whereas $l_{\alpha}(t)$ is attributed to an anomalous
behavior with a larger $z$.

If for the high temperature case the observed exponent $z=3$ is
regarded as the true dynamic behavior it is interesting to discuss how
such a value characteristic for a locally conserved order parameter
can be physically realized in our system. A suitable order parameter
for the solidification phase field would be the density difference
between ice and vapor. Nominating evaporation-condensation as the
dominant mechanism of mass transport does immediately imply a locally
conserved dynamics for the order parameter. It is rather the fact that
the evaporation-condensation dynamics is limited by the intermediate
diffusion which ultimately fixes $z=3$ \cite{lifshitz_1961}. If
instead the evaporation-condensation dynamics were limited by the
interface reaction one would end up with $z=2$
\cite{lifshitz_1961,wagner_1961}. In the reaction limited regime
diffusion can be regarded as infinitely fast, hence evaporated mass is
immediately available for condensation at \emph{remote} parts of the
structure only subject to a global mass conservation. This effectively
translates into an order parameter, which is only globally conserved
which still leads to $z=2$ \cite{lifshitz_1961}, similar to the
dynamics in the absence of any conservation law. The reaction limited
case requires a small condensation coefficient \cite{libbrecht_2005},
\cite{legagneux_2004} on the interface and thus our result $z=3$ is
consistent with diffusion-limited evaporation-condensation.

Despite the absence of dynamic scaling, all length scales discussed so
far still display a monotonic increase and can be reasonably well
described by a power law which is the same for all coordinate
directions. A completely different behavior is observed for the
structural correlation lengths $l_{0,\alpha}$, which is non-monotonic
and anisotropic. Remarkably, the observed anisotropy between the
rescaling properties of $C(r,t)$ in $z$ and $x$ direction
(Fig.~\ref{fig:6},\ref{fig:7}) has no influence on the evolution of
the interfacial correlation lengths $l_{\alpha}$ in the respective
directions, cf.~Fig.~\ref{fig:12}. This implies that the dynamics on
the smallest scales remains isotropic if prefactors are
neglected. This holds true at all temperatures and suggests that the
anisotropy emerges from larger scales. This is confirmed by the
evolution of the first zero crossing $l_{0,\alpha}$ of
$C_{\alpha}(r,t)/C(0,t)$ which displays a complicated, anisotropic
behavior. Usually, the zero crossing is interpreted as a typical
domain size. Hence the vector $\Omega(t)\sim(l_{0,x},l_{0,y},l_{0,z})$
can be interpreted as an oriented structural unit of the ice network
on scales larger than the previously discussed interfacial scale
$l_{\alpha}$. The dynamics of $\Omega(t)$ characterizes the evolution
of some orientational order which does not show any qualitative
differences between different temperatures, cf.~Fig.~\ref{fig:13}: The
$z$ component of $\Omega(t)$ always decreases rapidly followed by a
slow monotonic increase. The rapid decrease might be caused by some
rotational motion under gravity. This might be corroborated by the
fact that after one year the long arm of the dendrite is predominantly
found in the $x-y$ plane. In contrast the $x,y$ components of $\Omega$
show initially some transient behavior which is followed by a maximum
where correlations roughly double when compared to the initial
value. The temporal resolution is too coarse to investigate this
phenomenon in quantitative detail but the qualitative features of the
evolution are present at all temperatures. A quantitative
investigation of these structural rearrangements, their possible
\emph{impact} on modulations of the bulk densification rate and their
possible \emph{initiation} by interfacial scale coarsening is left for
future work.

Finally, we comment on the connection between the observed properties
of the correlation function and interfacial curvature distributions
which are often used to characterize the evolution of snow
\cite{flin_2003,legagneux_2004}. The small scale expansion of the
correlation function $C(r)$ for smooth but otherwise arbitrary
microstructures can be written as
$C(r)=(\phi-\phi^2)-(s/4)\,r-F[H,K]\,r^3 + \ldots$ in terms of the
volume fraction $\phi$, specific surface area $s$ and the mean and
Gaussian curvature fields $H(x)$ and $K(x)$, respectively (see
e.g.~\cite[Ch. 2]{torquato_2002}). Details which are irrelevant for
the present argument are subsumed in the functional $F$. The expansion
implies that the observed breakdown of dynamic scaling beyond the
interface scale $r=1/s$ is equivalent to the fact that the typical
curvatures $H,K$ have a different dynamic evolution as the specific
surface area $s$ and are more likely influenced by larger scale
structural re-arrangements. Hence, the absence of scaling for the
curvature distributions observed in \cite{legagneux_2004} is in
agreement with the absence of scaling in the correlation function at
larger scales. Our analysis above suggests that this is an implication
of the structural reorganization at larger scales and thus cannot be
predicted by coarsening mechanisms alone. Previous work on isothermal
metamorphism has usually focused on a single length scale $1/s$ and
its interpretation in terms of modified LSW approaches. This is
equivalent to neglect correlations of the density beyond the smallest
interfacial scales. This appears to be insufficient due to the
existence of different relevant length scales which are influenced by
memory effects of initial conditions or gravity. A first attempt to
include gravity in isothermal metamorphism has been done in
Monte-Carlo simulations \cite{vetter_2009}. Future effort in this
direction is highly desirable.

\section{Conclusions} \label{sec:6}

We have investigated isothermal metamorphism of snow as an example of
a material undergoing interfacial relaxations on the smallest scales
with concurrent structural relaxations on larger scales. If the
initial conditions contain density correlations on length scales which
are large compared to the interfacial correlation length strong memory
effects of the initial state can be expected. This appears to be one
of the reasons for the breakdown of dynamic scaling for coarsening of
snow.

The overall behavior of the interfacial relaxation is consistent with
coarsening of fractal clusters. The underlying dynamics is governed by
diffusion-limited evaporation-condensation with $z=3$ at high
temperatures and surface diffusion with $z=4$ at low temperatures. In
addition the dynamics at the scale of a typical, structural domain
size displays a rather complex, non-monotonic behavior. The latter is
attributed to structural relaxations of the ice network under the
influence of gravity. Apparently, the strong variations of the
dynamics on larger scales leave the interfacial phase ordering nearly
unaffected. Vice versa it poses the interesting question how large
scale, structural mobility of the system can effectively be induced by
interfacial coarsening.

\begin{acknowledgments}
  We gratefully acknowledge the valuable support of Matthias Jaggi and
  Stephen Steiner during the experiments in the cold lab.
\end{acknowledgments}


\end{document}